\documentclass[preprint,aps,floatfix,unsortedaddress]{revtex4}
\usepackage{graphicx}
\usepackage{amsmath}
\usepackage{longtable,array}
\usepackage{lscape}
\usepackage{color}
\usepackage{multirow}
\usepackage{soul}
\usepackage{threeparttable}
\usepackage{rotating}
\usepackage{epsfig}

\newcommand{\etal}{{\it et al.} }
\newcommand{\ai}{{\it ab initio}}

\newcommand{\cm}{cm$^{-1}$}

\newcommand{\sothree}{$^{32}$S$^{16}$O$_3$}
\newcommand{\pmax}{$P_{\rm max}$}
\newcommand{\angstrom}{\mbox{\normalfont\AA}}

\newcommand{\3}{$_{3}$}

\newcommand{\Dh}{${\mathcal D}_{3{\rm h}}$(M)} 

\newcommand{\p}{^\prime}
\newcommand{\pp}{^{\prime\prime}}

\def\a0{{$a_{\rm 0}$}}

\begin{document}

\title{An \ai\ variationally computed room-temperature line list for \sothree }

\author{Daniel S. Underwood, Jonathan Tennyson and Sergei N. Yurchenko}
\email[To whom correspondence should be addressed: ]{s.yurchenko@ucl.ac.uk}

\affiliation{Department of Physics and Astronomy, University College London,
Gower Street, London WC1E 6BT, United Kingdom}



\date{\today}

\begin{abstract}
{\it Ab initio}  potential energy and dipole moment surfaces are computed
for sulfur trioxide (SO$_3$) at the CCSD(T)-F12b level of theory and appropriate triple-zeta basis sets.
These surfaces are fitted and used, with a slight correction for the
equilibrium S--O distance, to compute pure rotational and rotation-vibraton spectra of \sothree\
using the variational nuclear motion program TROVE.
The calculations considered transitions in the region 0--4000 \cm\ with rotational states up to $J=85$.
The resulting line list of 349~348~513 transitions is appropriate for modelling room temperature \sothree\ spectra.
Good agreement is found with the observed infrared absorption spectra and
the calculations are used to place the measured relative intensities on an absolute
scale. A list of 10~878 experimental transitions is provided in a form suitable for
inclusion in standard atmospheric and planetary spectroscopic databases.

\end{abstract}

\maketitle

\section{Introduction}
\label{s:intro}

Sulfur trioxide (SO$_3$) is a stable, planar, symmetric molecule whose
electronic ground state is a closed shell. On earth it is a pollutant
produced from smoke-stacks and other industrial exhausts
\cite{05RaHeSoOa.SO3}. In the atmosphere SO$_3$ forms sulfuric acid
with its association with acid rain, and inside combustion systems it
is corrosive. In both cases the reactivity of SO$_3$ makes it
production hard to monitor \cite{12FlVaAnBr.SO3}. SO$_3$ is produced
naturally on earth as part of volcanic emissions \cite{05MiKrGrAn.SO3}.
It is also thought to be a significant constituent of the atmosphere
of Venus \cite{10ZhLiMoBe.SO3}.

The infrared vibration-rotation spectrum of \sothree\ (henceforth referred to as SO\3) has been extensively  investigated
in a series of papers by Maki and co-workers \cite{73KaMaDo.SO3, 89OrEsMa.SO3,01ChVuMa.SO3,01MaBlSa.SO3,02BaChMa.SO3,03ShBlSa.SO3,04MaBlSa.SO3}. Its
``forbidden'' rotational spectrum, for which centrifugal distortions can
induce transitions, has been investigated using microwave Fourier-transform
spectroscopy \cite{91MeSuDr.SO3}. However, although these works provide
an extensive list of measured line frequencies, none of them report
absolute transition intensities.

There has been limited theoretical work on SO$_3$. Dorney \etal\ \cite{73DoHoMi.SO3} reported force constants, while
Martin~\cite{99Maxxxx.SO3} computed an \ai\ potential energy surface (PES)
using coupled cluster methods and reported theoretical estimates for
the band origins of the low-lying vibrational states. Again, neither of
these works considered transition intensities.

The lack of any absolute transition intensities for SO$_3$ places severe
limitations on the use of its infrared spectrum for remote sensing applications
or inclusion of this data in standard atmospheric and planetary
spectroscopic databases \cite{jt453,jt504}. In this work we report
the calculation of a new \ai\ PES and associated
dipole moment surfaces (DMS) for SO\3. These are used to not only
produce theoretical spectra for \sothree, but also to place
the relative intensity measurements of its infrared spectrum on an absolute
scale.


\section{The potential energy and dipole moment surfaces}
\label{s:PES-DMS}

The \ai\ PES was computed using the recently-proposed explicitly
correlated F12 singles and doubles coupled cluster method including a
perturbational estimate of connected triple excitations,
CCSD(T)-F12b~\cite{Adler07}, in conjunction with the corresponding
F12-optimized correlation consistent basis sets, namely the valence
correlation-consistent functions aug-cc-pVTZ-F12 and
aug-cc-pV(T+d)Z-F12 for oxygen and sulfur,
respectively~\cite{Yousaf09}. We also utilized the
OptRI~\cite{Yousaf08} cc-pV5Z/JKFIT~\cite{Weigend02} and
aug-cc-pwCV5Z/MP2FIT~\cite{Haettig05} auxiliary basis sets for
evaluating the many-electron integrals, the exchange and Fock
operators, and the remaining electron repulsion integrals,
respectively. The value of the geminal Slater exponent $\beta$ was
chosen as 1.2.
In \ai\ calculations Molpro.2010 \cite{molpro} was employed.
We used a large
grid of 13~000 points with the stretching modes ranging from 1.1 to 2.0
\AA, interbond angles $\alpha$  between  70 and 123$^\circ$, and the inversion
angle $\rho$ between 0 and 50$^\circ$. This grid was sufficient to cover
energies up to 40~000~\cm\ above the minimum.

This PES was then
represented in an analytical form given by the
expansion~\cite{05YuCaJe.NH3}
\begin{eqnarray}
V(\xi_1,\xi_2,\xi_3,\xi_{4a},\xi_{4b};{\sin\bar \rho}) \,
    &=&  V_{\rm e} +  V_0({\sin\bar \rho}) + \sum_j \, F_j({\sin\bar \rho}) \, \xi_j
\nonumber \\
& + &
       \sum_{j \leq  k} \, F_{jk}({\sin\bar \rho}) \, \xi_j \, \xi_k
       +  \sum_{j \leq k \leq l} \, F_{jkl}({\sin\bar \rho}) \,\xi_j \, \xi_k \, \xi_l
\nonumber \\
       &+ & \sum_{j \leq k \leq l \leq m} \, F_{jklm}({\sin\bar \rho}) \, \xi_j \, \xi_k \,
        \xi_l \, \xi_m
\label{e:PEF:morbid}
\end{eqnarray}
in the coordinates $\xi_k$:
\begin{eqnarray}
\label{CurvStretc1}
   \xi_{k\phantom{a}}  &=& 1-\exp( -a( r_k -r_{\rm e} ) ) \hbox{,} \;\; k=1,2,3, \\
   \label{CurvXi4a}
   \xi_{4a}            &=& \frac{1}{\sqrt{6}} \left( 2 \alpha_{1} - \alpha_{2} - \alpha_{3} \right)\hbox{,} \\
   \label{CurvXi4b}
   \xi_{4b}            &=& \frac{1}{\sqrt{2}} \left(\alpha_{2} -  \alpha_{3} \right)\hbox{,} \\
   \label{e:sin:brho}
   \sin {\bar \rho} \, &=& \, \frac{2}{\sqrt{3}} \,\sin [
     (\alpha_{1} + \alpha_{2} + \alpha_{3}) /6] \hbox{,}
\end{eqnarray}
where
\begin{equation}
\label{V0}
V_0(\sin {\bar \rho}) = \sum_{s=1} \, f_0^{(s)} \, (\sin \rho_e - \sin {\bar
  \rho})^s
\end{equation}
and
\begin{equation}
F_{jk\dots}(\sin{\bar \rho}) = \sum_{s=0} \, f_{jk\dots}^{(s)} \, (\sin \rho_e -
\sin {\bar \rho})^s  .
\end{equation}
The same form has been used to  represent the PESs of NH\3,
PH\3, SbH\3, NH\3$^{+}$, and BiH\3
\cite{02LiThYu.NH3,06YuThJe.SbH3,08OvThYu.PH3,08YuThCa.NH3+}.
The potential parameters $f_{jk\ldots}^{(s)}$ were obtained through a
least squares fit to the \ai\ points with an root-mean-squares (rms)
error of 0.067 \cm. Weight factors were set using the expression
suggested by Partridge and
Schwenke~\cite{ps97}:
\begin{equation}
\label{e:weights}
w_i = \frac{\tanh \left[ -0.0005\;\mbox{cm} \times \left(V_i - 16\,000\;\mbox{cm}^{-1}\right)\right] + 1.002\,002\,002}
{ 2.002\,002\,002\;{\rm cm} \times V_i^{(w)}},
\end{equation}
where $V_i^{(w)}= \max(16\,000\;\mbox{cm}^{-1},V_i)$, and $V_i$ is the
\ai\ energy at the $i$th geometry (in \cm), measured relative to the
equilibrium energy. The \ai\ energy $V_i$ is weighted by the factor
$w_i$ in the PES fitting; these weight factors favor the energies
below 16~000 cm$^{-1}$.  The \ai\ equilibrium geometry $r_{\rm e}$ as
obtained from the fitting is 1.42039~\AA. This can be compared with an
experimentally derived value 1.41732 \angstrom ~\cite{89OrEsMa.SO3}.
We found that the experimental rotational
energies of SO\3\ ~\cite{01MaBlSa.SO3} are best
described by the latter value. Therefore we decided to use this value of
$r_{\rm e}$ in all our calculations in place of the \ai\ value. It is
known however that replacing the equilibrium structure may cause
undesirably large changes to the ro-vibrational energies. In order to
minimize this effect the following procedure was employed.

The idea is to expand the \ai\ PES around some reference value $r_{\rm
  ref}$ and than remove the linear terms in the PES expansion and
replace the expansion center with $r_{\rm e}^{\rm (exp)}$. In doing
this we expect the change of the shape in the PES to partly compensate the
effect from  replacing $r_{\rm e}^{\rm (ai)}$ with $r_{\rm e}^{\rm
  (exp)}$ on the vibrational  energies.
We choose the reference center $r_{\rm e}^{\rm (ref)} = 1.42039 + \Delta r_{\rm e}$,  where
$\Delta r_{\rm e} = r_{\rm e}^{\rm (ai)}-r_{\rm e}^{\rm (exp)} = 0.00307$, i.e. on
the opposite side from $r_{\rm e}^{\rm (exp)}$  and at the same distance from $r_{\rm e}^{(\rm exp)}$.

This is the only adjustment to the shape of \ai\ PES utilized. Table~\ref{rotationComp} compares rotational levels obtained by Maki \etal~\cite{01MaBlSa.SO3} and those computed using
the equilibrium-adjusted PES used in this work. The agreement is very good.

The DMS were calculated using the same level of theory as the PES and
on the same grid of 13~000 points. The \ai\ values were then expressed
analytically using the symmetrized molecular bond (SMB) representation described in detail in Ref.~\cite{jt500}.
The resulting dipole moment parameters obtained through a least squares fit reproduce the \ai\
data with an rms error of 0.00013~D. In these fittings the same factors
defined by Eq.~\eqref{e:weights} were used to weight the geometries
according to the corresponding energies. To our knowledge there are no
experimental or \ai\ dipole moment data in the literature that
we could use to validate our DMS against.  However,
our experience of working dipole moments for different systems
\cite{jt156,jt500,jt509,13YuTeBa.CH4} shows that  \ai\
intensities in most cases are competitive with experimental
measurements. The quality of the relative intensities calculated using
our DMS is discussed below.

Both the potential energy and dipole moments functions used in the present work are given as supplementary data.


\begin{center}
\begin{table}[ht]
\caption{Theoretical rotational term values for \sothree\ ground
vibrational state (\cm) compared with experiment~\cite{01MaBlSa.SO3}.}
\centering
\scriptsize \tabcolsep=12pt
\begin{tabular}{rrrr}
\hline\hline
$J$ & $K$ & Obs. & TROVE \\
\hline
 2 & 0  & 2.0912 & 2.0916\\
 3 & 3  & 2.6115 & 2.6119\\
 4 & 3  & 5.3998 & 5.4006\\
 4 & 0  & 6.9707 & 6.9718\\
 5 & 3  & 8.8852 & 8.8864\\
 6 & 6  & 8.3548 & 8.3559\\
 6 & 3  & 13.0675 & 13.0694\\
 7 & 6  & 13.2342 & 13.2360\\
 7 & 3  & 17.9467 & 17.9493\\
 8 & 6  & 18.8106 & 18.8132\\
 8 & 3  & 23.5228 & 23.5263\\
 8 & 0  & 25.0935 & 25.0972\\
 9 & 9  & 17.2297 & 17.2319\\
 9 & 6  & 25.0838 & 25.0874\\
 9 & 3  & 29.7958 & 29.8002\\
10 & 10 & 17.2297 & 17.2319\\
10 & 9 & 24.2002 & 24.2035\\
10 & 6 & 32.0539 & 32.0584\\
10 & 3 & 36.7655 & 36.7709\\
10 & 0 & 38.3360 & 38.3417\\
20 & 18 & 89.8252 & 89.8372\\
20 & 15  & 107.0973 & 107.1122\\
20 & 12  & 121.2253 & 121.2425\\
20 & 9  & 132.2114 & 132.2304\\
20 & 6  & 140.0574 & 140.0777\\
20 & 3  & 144.7645 & 144.7856\\
20 & 0  & 146.3334 & 146.3548\\
80 & 78 & 1195.6589 & 1195.8085\\
\hline\hline
\end{tabular}
\label{rotationComp}
\end{table}
\end{center}

\section{Ro-vibrational calculations}

\subsection{Basis set convergence and Hamiltonian optimisation}

Ro-vibrational calculations were performed with the program TROVE \cite{07YuThJe.method} adapted to work in the
\Dh\ molecular permutation-inversion
group appropriate for SO\3.
In order to achieve results of high accuracy as well as minimising the
requirement  for computational  resources, it is  necessary to optimise
the  size  of the   Hamiltonian matrix.  This  involves   preliminary
truncation of the basis set, as well as limiting the order of both the
kinetic and potential components of  the Hamiltonian expansion.  TROVE
employs  a polyad number  truncation which  controls  the size  of the
basis set. For SO\3\ the polyad number is given by
\begin{equation}
\label{e:polyad}
P = 2(n_1 + n_2 + n_3) + n_4 + n_5 + \frac{n_6}{2},
\end{equation}
where $n_i$ are the quanta associated with 1D basis functions, $\phi_i$,
whose product gives our vibrational basis set \cite{07YuThJe.method}.
Each of these basis functions is associated with an internal coordinate
$ \xi_{i}$, and only functions for which $P \le$ \pmax\ are included in
the primitive basis set. Initial tests were carried out to measure the
degree of convergence using different values for \pmax, and the orders
of the kinetic and potential energy expansions. In this work we use a
kinetic energy expansion of order 4, and a potential energy expansion
of order 8; using a kinetic energy expansion order of 6 requires a
more expensive calculation where convergence is already observed to
within 0.001 \cm\ when expanding to fourth order.  In the present
study, we find that the convergence is more sensitive to \pmax, and we
obtain convergence to within 0.1 \cm\ when \pmax\ is 12 or 14
(see table ~\ref{fundamentals}), therefore we use a basis set based on
\pmax = 12.

\begin{table}[ht]
\caption{Convergence of basis set viewed for some vibrational band centres (\cm) for \sothree.}
\centering
\begin{tabular}{lrrrr}
\hline\hline
& Obs.~\cite{01MaBlSa.SO3} &  \pmax = 10 & \pmax = 12 & \pmax = 14\\
\hline
$\nu_1$ & 1064.92 & 1065.83 & 1065.75 & 1065.74\\
$\nu_2$ & 497.57 & 498.48 & 498.48 & 498.48\\
$\nu_3$ & 1391.52 & 1387.63 & 1387.45 & 1387.43\\
$\nu_4$ & 530.09 & 528.61 & 528.59 & 528.58\\
\hline
2$\nu_3$($l_{3}$=0) & 2766.40 & 2759.61 & 2759.12 & 2758.75\\
2$\nu_3$($l_{3}$=2) & 2777.87 & 2770.70 & 2770.29 & 2769.95\\
\hline
2$\nu_2$ & 995.02 & 995.43 & 995.35 & 995.35\\
2$\nu_4$($l_{4}$=0) & 1059.81 & 1057.10 & 1056.50 & 1056.44\\
2$\nu_4$($l_{4}$=2) & 1060.45 & 1057.86 & 1057.38 & 1057.33\\
$\nu_2$ + $\nu_4$($l_{4}$=1) & 1027.90 & 1027.58 & 1027.35 & 1027.33\\
\hline
$\nu_1$ + $\nu_4$($l_{4}$=1) & 1593.69 & 1593.82 & 1593.36 & 1593.30\\
3$\nu_4$($l_{4}$=1) & 1589.81 & 1587.64 & 1586.46 & 1586.30\\
3$\nu_4$($l_{4}$=3) & 1591.10 & 1587.61 & 1586.43 & 1586.27\\
$\nu_1$ + $\nu_2$ & 1560.60 & 1565.51 & 1565.33 & 1565.32\\
$\nu_2$ + 2$\nu_4$($l_{4}$=0) & 1557.88 & 1556.38 & 1555.59 & 1555.47\\
$\nu_2$ + 2$\nu_4$($l_{4}$=2) & 1558.52 & 1557.12 & 1556.45 & 1556.37\\
2$\nu_2$ + $\nu_4$($l_{4}$=1) & 1525.61 & 1524.81 & 1524.48 & 1524.46\\
3$\nu_2$ & 1492.35 & 1449.81 & 1490.76 & 1490.76\\

$\nu_2$ + $\nu_3$($l_{3}$=1) & 1884.57 & 1881.82 & 1881.53 & 1881.51\\
3$\nu_3$($l_{3}$=1) & 4136.39 & 4138.88 & 4126.78 & 4125.92\\
\hline\hline

\end{tabular}
\label{fundamentals}
\end{table}

As well as using the polyad number to truncate the size of the basis,
we employ a further truncation technique by specifying an upper limit for the
eigenvalue calculations, i.e. construct the basis set such that it
provides energy values up to a limit of $E_{\rm max}$. This is based on an
estimation whereby eigenvalues of our 1D basis functions are summed
together before they are considered for matrix element calculations,
and the active space is constructed using basis functions whose
eigenvalues sum together to have $E \le E_{\rm max}$. For the present
study we use $E_{\rm  max}/hc$ = 10~000 \cm.

These precautions are particularly important for the SO\3\
molecule, as its larger mass (compared to, for example, XH$_3$
systems) gives rise to small rotational constants, which in turn
requires calculations up to high $J$ value to ensure adequate coverage
of transitions for a given temperature. This means that any
unnecessary basis functions will prove computationally expensive. In
addition to this basis set minimisation, we can reduce the size of the
Hamiltonian further by making use of group theory. SO\3\
has \Dh\ molecular group symmetry and  the spin-0 Bosons
which make up the constituent atoms of this molecule allow the molecular
ro-vibrational wavefunctions to exhibit the symmetry of only two of the
six irreducible representations of this group in order to satisfy the
Pauli Principle; namely the A$_1\p$ and A$_1\pp$ representations.
This reduces both the number of Hamiltonian matrices we need to consider and,
since $E$ symmetry Hamiltonian matrices are larger, their size.

\begin{figure}[ht]
\centering
\scalebox{0.4}{\includegraphics[angle=270]{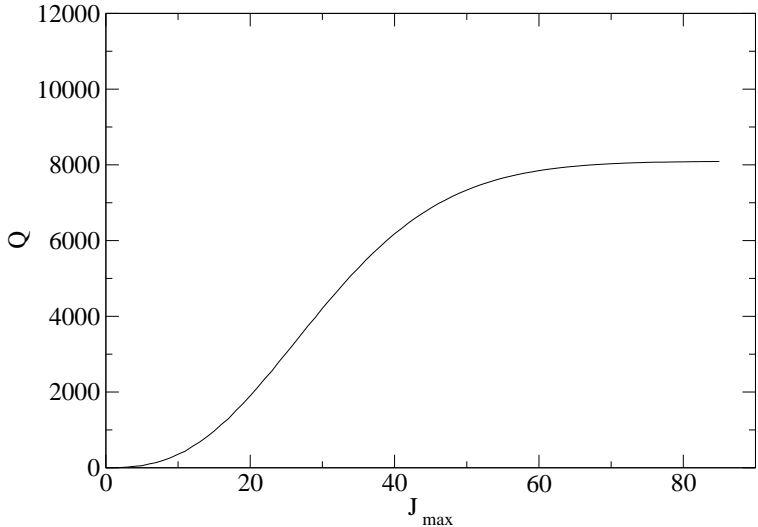}}
\caption{Convergence of partition function for different values of $J_{\rm max}$.}
\label{pFuncConv}
\end{figure}

The two other factors which are important in our spectral calculations are
i) the wavenumber range of the desired synthetic spectrum, and ii) the
temperature at which we wish to simulate it. The quality of a computed
spectrum will become sensitive to $E_{\rm max}$ as the temperature
increases; we need to ensure that we calculate all energy states that
are significantly populated for the given temperature.  This can be
checked using the temperature-dependent partition function:
\begin{equation}
\label{e:pf}
    Q = \sum_{i}\ g_{i} \exp(-E_{i}/kT),
\end{equation}
where $g_{i}$ is the total degeneracy of the ro-vibrational state $i$
with energy $E_{i}$, with the sum running over all energies at the
absolute temperature $T$, and $k$ is Boltzmann's constant. The total
degeneracy is given by $(2J + 1)$ times the nuclear spin degeneracy,
which for the present case of \sothree\ is simply 1 for both the
A$_1\p$ and A$_1\pp$ symmetries, given that the nuclear spin of  $^{16}$O and $^{32}$S are zero.
For a given temperature, we can determine the
contribution of various states to the value of $Q$. We can then check
that $Q$ converges to a specific value as $E_{i}$ tends to infinity;
as $T$ increases we require a greater coverage of higher-lying energy
states. For $T$ = 298.15~K we find that $Q$ converges to better than 1~\%\ at $J = 85$, with a value of $Q = 8089.262$.  Therefore calculations
spanning all $J$'s up to 85 should be sufficient for
simulating spectra at this temperature. Figure ~\ref{pFuncConv} shows
the value of $Q$ as a function of all energy levels having $J$ quantum
number up to a maximum value, $J_{\rm max}$, at an absolute
temperature of $T$ = 298.15 K. As we include energies in the summation
for increasing values of $J$ we see that the associated energy levels
contribute less and less to the value of $Q$, until it converges to a
limit.

An \ai\ PES was previously published by Martin based on coupled cluster methods~\cite{99Maxxxx.SO3}. As an intial test of the TROVE procedure we used the quartic force field by Martin to compute fundamental term values, and to test our convergence procedures, using a polyad truncation scheme of \pmax = 16. We found discrepancies between our results and the values published, particularly in the value of the $\nu_2$ fundamental term value. We made a substitution to the symmetry-adapted force constant $F_{22}$ associated with this vibration, taking a scaled value from a previously published force-field~\cite{73DoHoMi.SO3}, which resulted in a fair improvement. Table \ref{martinComp} illustrates the differences between the values and our final converged result for the fundamental vibrations.

\begin{table}[ht]
\caption{A comparison of the fundamental term values (\cm) between this work and Martin~\cite{99Maxxxx.SO3} (computed with TROVE).}
\centering
\begin{tabular}{c rrrr}
\hline\hline
& Obs. ~\cite{01MaBlSa.SO3} & Martin & Martin $F_{22}$\tnote{a} & This work\\
\hline
$\nu_1$ & 1064.92 & 1063.36 & 1064.22 & 1065.75\\
$\nu_2$ & 497.57 & 428.36 & 487.10 & 498.48\\
$\nu_3$ & 1391.52 & 1386.81 & 1386.85 & 1387.45\\
$\nu_4$ & 530.09 & 527.35 & 527.32 & 528.59\\
\hline
\hline
\end{tabular}
\begin{tablenotes}
\item[a] Martin, with a substituted  $F_{22}$ value ~\cite{73DoHoMi.SO3}.
\end{tablenotes}
\label{martinComp}
\end{table}

\subsection{Intensity simulations}

The simulation of absorption spectra can be broken down into two main
parts: first, the eigenfunctions and eigenvalues of the numerically
constructed Hamiltonian matrix are calculated by a diagonalisation
procedure; second, these eigenfunctions are used to compute transition
dipoles, line strengthes $S$, Einstein $A$ coefficients and intensities $I$ for allowed transitions.
For \sothree, the rigorous selection rules determining
allowed electric dipole transitions are $\Delta J = J\p - J\pp = 0, \pm 1 \
(J\pp + J\p \ge 1)$, and symmetry selection rules $A_1\p \leftrightarrow A_1\pp$.
The intensity of a transition between given states is given by the formula
\begin{equation}
\label{e:Intensity} I(f \leftarrow i) =
\frac{8\pi^3 N_{A}{\nu_{if}}}{(4\pi \epsilon_{0}) 3hc}\ \frac{e^{-E_i/kT}}{Q(T)}\ \times\
\left[1 - \exp\left(\frac{-hc\nu_{if}}{kT}\right) \right]\ S(f \leftarrow i),
\end{equation}
where  $I(f \leftarrow i)$ is the transition intensity for a transition from state $i$
with energy $E_i$ to state $f$ with energy $E_f$, with $hc\nu_{if} = E_f - E_i$. $Q$ is the partition function.
The value of $Q$ must be converged for the absolute temperature $T$, and with respect to the wavenumber range
in which we wish to simulate our spectrum.
$S(f \leftarrow i)$ is the line strength, which is defined by the following integration:
\begin{equation}
\label{e:lineStrength} S(f \leftarrow i) = g_{\rm ns} \sum_{m_f, m_i} \sum_{X,Y,Z} |\langle \Phi_{\rm rv}^{(f)}|\mu_{A}|\Phi_{\rm rv}^{(i)}\rangle|^2
\end{equation}
for a transition between initial state $i$ with rovibrational wavefunction $\Phi_{\rm rv}^{(i)}$ and final state
$f$ with wavefunction $\Phi_{\rm rv}^{(f)}$. Here, $g_{\rm ns}$ is the nuclear spin statistical weight factor, and $\mu_{A}$
is the electronically averaged component of the molecular dipole moment along the space-fixed axis $A = X, Y, Z$.
The quantum numbers $m_i$ and $m_f$ are projections of the total angular momentum $J$ (in units of $\hbar$) on the laboratory fixed
$Z$ axis, for the initial and final states, respectively.

Maki \etal \cite{73KaMaDo.SO3,89OrEsMa.SO3,01ChVuMa.SO3,01MaBlSa.SO3,02BaChMa.SO3,03ShBlSa.SO3,04MaBlSa.SO3}
reported an extensive high-resolution study of a number of fundamental, combination
and overtone bands of \sothree. 
Their principle aim was to obtain accurate wavenumber measurements, but
relative intensities were also measured. In the present work we convert these data into absolute intensities by normalizing to the theoretical intensities obtained with TROVE at $T$ = 298.15~K as described below.

The measurements available to us cover three spectral regions: 405--708~\cm\ (focusing on $\nu_2$, $\nu_4$, 2$\nu_2$-$\nu_2$, $\nu_2$+$\nu_4$-$\nu_2$, $\nu_2$+$\nu_4$-$\nu_4$,  $\nu_1$-$\nu_4$, 2$\nu_4^{(l_4=0)}$-$\nu_4$ and $2\nu_4^{(l_4=2)}-\nu_4$), 1200--1680~\cm ($\nu_3$), and 2500-3280~\cm\ ($2\nu_3^{(l_3=2)}$). Each measurement was made at different values of pressure. For the 405--708 \cm\ window, the measurements were performed at 0.409 and 2.04 Torr, and for the $\nu_3$ measurements
between the 1200--1680 \cm\ window 0.16 Torr and 0.7 Torr was used. 560 lines were measured at 0.7 Torr, however 439 of these had relative intensity values which were negative. We therefore did not use this higher pressure measurement at all.

In Table ~\ref{numLines} we compare the numbers of lines identified in each measurement to the numbers of lines computed using TROVE. The latter numbers are
the subject of the following selection criteria:  $J\le 85$,  intensity cut-off,  $I(f \leftarrow i) > 10^{-34}$ cm/molecule, and the wavenumber window, 0--4000 \cm. Experimental lines with negative relative intensities were left out of the analysis.

To normalize the experimental intensities, the experimental relative data from each spectral window and each different pressure  were scaled to match the theoretical values computed at $T$ = 298.15~K. The scaling factors obtained through a minimization procedure using all selected experimental lines with non-zero intensity (see Table~\ref{bandInts}) are 6.571$\times 10^{-21}$ cm/molecule (405--708 \cm, 0.409 Torr), 1.838$\times 10^{-21}$ cm/molecule (405--708 \cm, 2.04 Torr), 4.823$\times 10^{-20}$ cm/molecule ( 1200--1680 \cm, 0.16 Torr), and 1.328$\times 10^{-21}$ cm/molecule (2500-3280 \cm, 4.99 Torr).
With these factors the `experimental' intensities match the theoretical values reasonably good, for example for 405--708~\cm\ the agreement is within about $7.02\times\ 10^{-22}$~cm/molecule and $5.05\times\ 10^{-22}$ cm/molecule, at 0.409 and  2.04 Torr, respectively.

Having the absolute intensities derived, band intensities were estimated for each experimental band as the sum of individual line intensities.  In Table~\ref{bandInts} these `experimental' band intensities $S^{\rm exp}$  (cm/molecule) are compared to the theoretical values obtain by summing intensities (a) from all TROVE lines  from a given window and (b) only from lines with experimental counterparts present. This was done separately for each spectral range, and each measurement pressure therein.  In Table \ref{bandInts} these quantities are referenced to as $S_{\rm tot}^{\rm calc}$ and $S_{\rm red}^{\rm calc}$  for the `total' and `reduced' band intensities, respectively, and compared to $S^{\rm exp}$.
The ratio $S_{\rm red}^{\rm calc}$ to $S^{\rm exp}$ also shown in Table~\ref{bandInts} demonstrates the good quality of the procedure employed as well as of our dipole moment. For example at 0.409 Torr, the differences between `experimental' and theoretical band intensities are within about $~$20\% for all  bands from the 405--708~\cm\ region with the exception of $2\nu_4^{(l_4 = 0)}$ - $\nu_4$ (see also discussion below). It should be stressed here that only one scaling factor for all eight bands from this window was applied at a given pressure. The difference between two theoretical band intensities $S_{\rm tot}^{\rm calc}$ and  $S_{\rm red}^{\rm calc}$  gives a measure of the missing experimental transition data. According to Table~\ref{bandInts} even stronger bands miss more that 50\% of the total intensity.

In Table~\ref{bandInts} we also show theoretical values of vibrational transition moments defined as
\begin{equation}\label{e:tm}
   \bar{\mu} =  \sqrt{ \bar{\mu}_{x}^2+\bar{\mu}_{y}^2+\bar{\mu}_{z}^2 },
\end{equation}
where
\begin{equation}\label{e:mu:me}
   \bar{\mu}_{\alpha} =  \langle \Psi_{\rm vib}^{(i)} \vert \mu_{\alpha}  \vert \Psi_{\rm vib}^{(f)} \rangle,
\end{equation}
and $\Psi_{\rm vib}^{(i)}$ and $\Psi_{\rm vib}^{(f)}$ are the  vibrational eigenfunctions of the `initial' and `final' states, respectively, variationally computed using TROVE and $\mu_{\alpha}$ is the component of the molecular dipole moment along the molecular-fixed axis $\alpha = x, y, z$.

\begin{figure}[t]
\begin{center}
{\leavevmode \epsfxsize=11.0cm \epsfbox{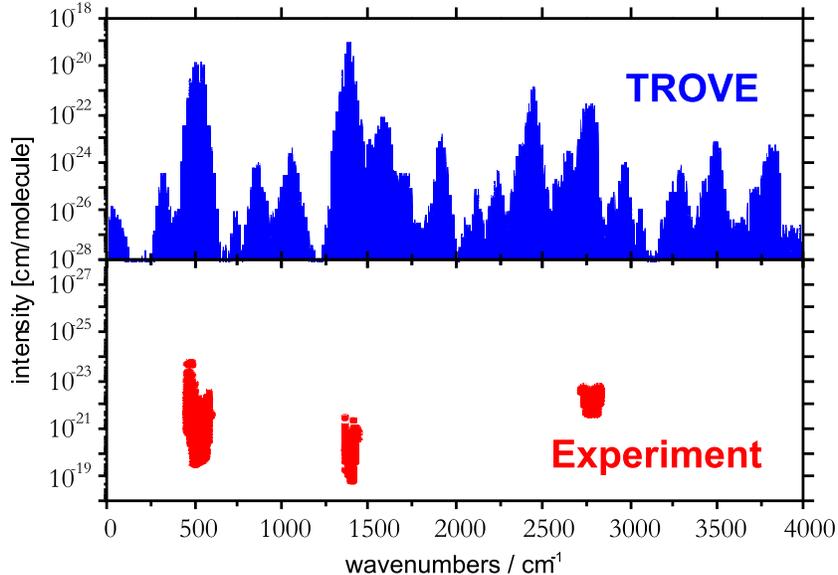}}
\end{center}
\caption{\label{f:overview} Overview of the simulated absorption ($T$ = 298.15K) spectrum (TROVE) of SO\3\ compared
to experiment scaled to the theoretical intensities (see text).}
\end{figure}

\begin{figure}[ht]
\centering
\scalebox{0.5}{\includegraphics[angle=270]{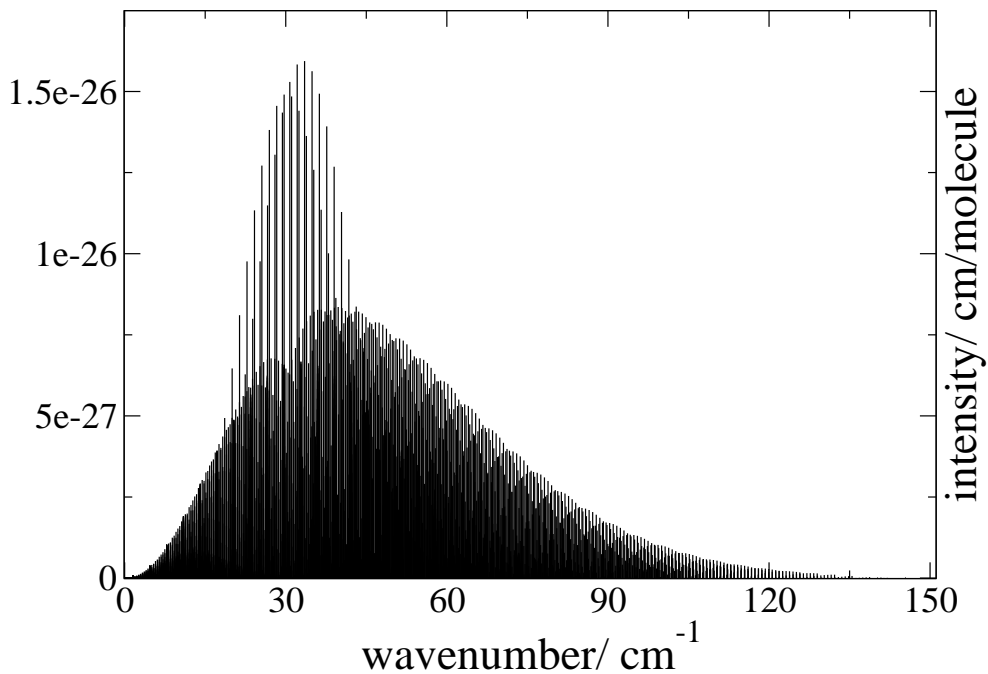}}
\caption{Rotational absorption band computed for $T$ = 298.15K, complete up to $J$=85.}
\label{micro}
\end{figure}

\begin{figure}[ht]
\centering
\scalebox{0.5}{\includegraphics{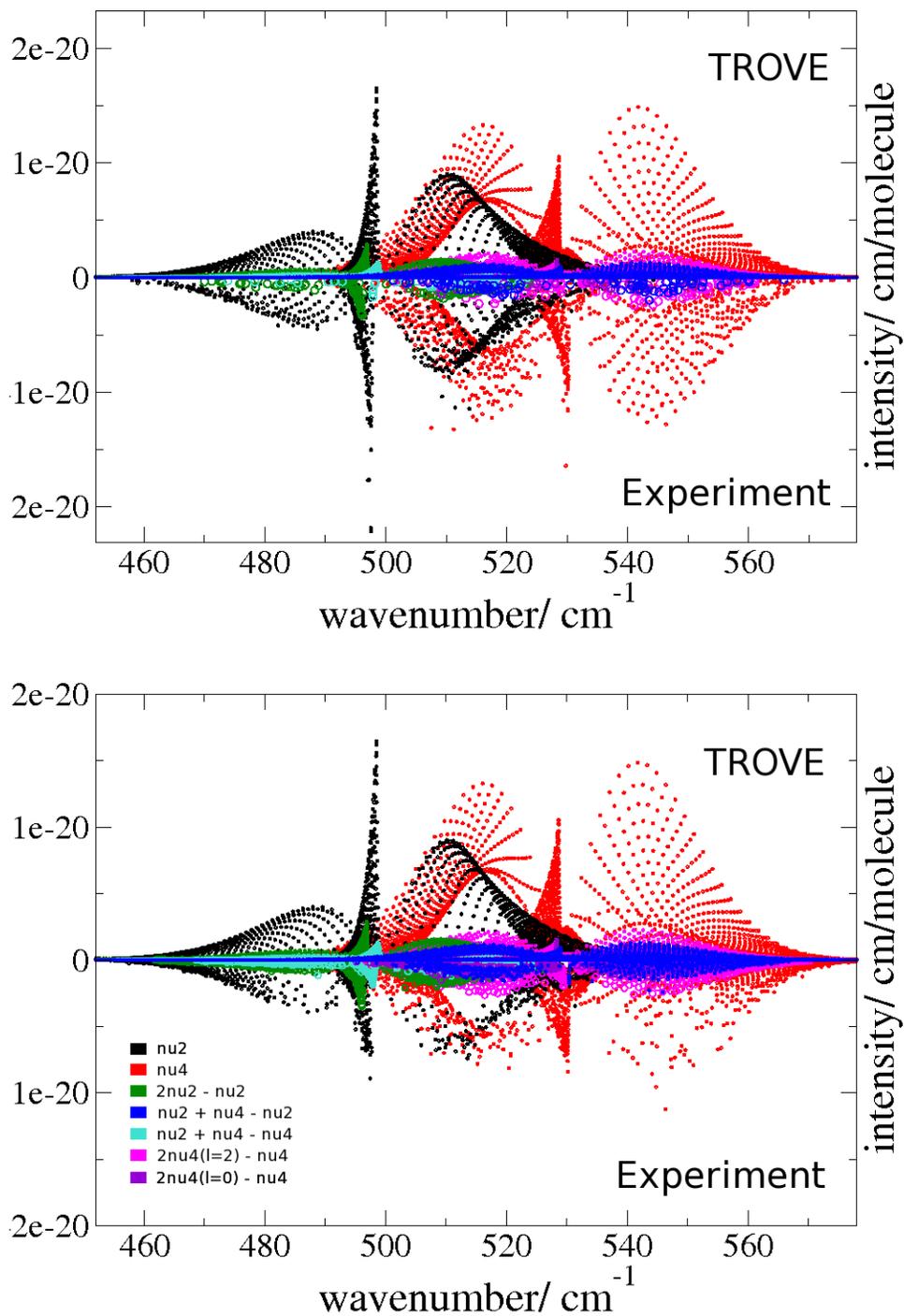}}
\caption{Comparison plot of TROVE results and the bands of interest measured between 405 - 708 \cm\ by Maki \etal,
 at 0.409 Torr (above) and 2.04 Torr (below). Points are enlarged in some cases for clarity.}
\label{allbands500}
\end{figure}

\begin{figure}[ht]
\scalebox{0.3}{\includegraphics{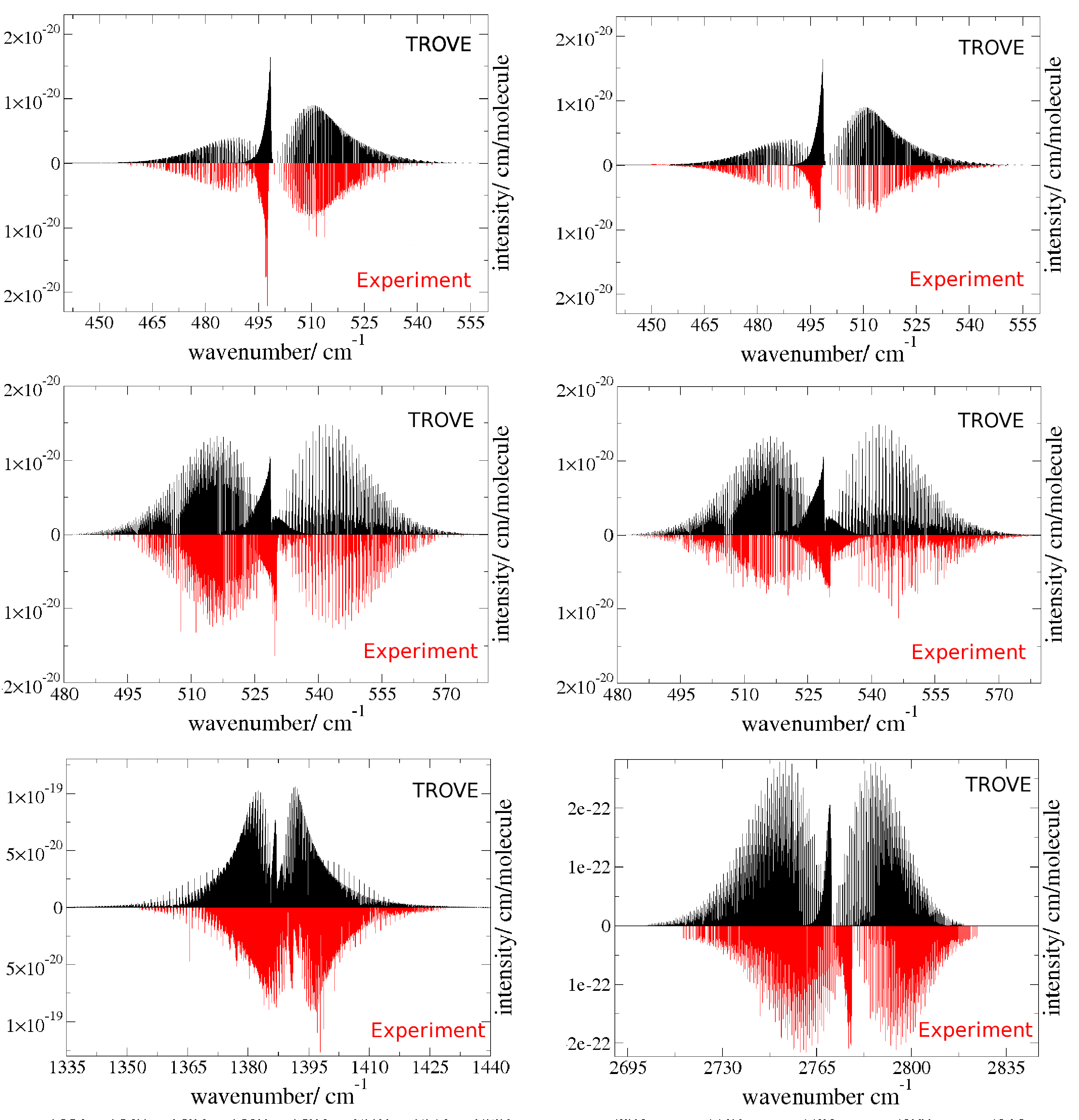}}
\caption{Fundamental band comparisons between this work ($T$ = 298.15K, complete up to $J$=85) and Maki \etal~\cite{01MaBlSa.SO3}.
The top and middle panels show comparisons for the $\nu_2$ and $\nu_4$ bands, respectively, at 0.409 Torr
(left) and 2.04 Torr (right).
The bottom panel shows a comparison for the $\nu_3$ band at 0.16 Torr (left) and the $2\nu_3$ band at 4.99 Torr (right).}
\label{so3fundamentals}
\end{figure}

Figure~\ref{f:overview} presents an overview of the simulated spectrum ($T$ = 298.15~K) with TROVE and experimental absorption spectra of SO\3\
 for the whole simulation range up to 4000~\cm. It reveals the gaps and limitations of the available experimental data.
Our intensities based on the \ai\ DMS are in very good qualitative agreement with experiment.
Figure~\ref{micro} shows the `forbidden' rotational band as a stick spectrum. It should be noted that the microwave measurements from Ref.~\cite{91MeSuDr.SO3}
 do not have any intensities reported.
In Figure ~\ref{allbands500} all eight bands from the 405--708 \cm\ region are combined into one graph for each pressure to illustrate the quality
 of the corresponding experimental data. This figure suggests that the 0.409 Torr data are generally more reliable. This is also reflected by the
 ratio values $S_{\rm red}^{\rm calc}/S^{\rm exp}$ from Table~\ref{bandInts}, which are significantly closer to 1 at the lower pressure.
Of the data present for two conditions we therefore place preference on scaled intensity values obtained at the lower pressure.
Finally, Figure ~\ref{so3fundamentals} presents a detailed comparison for the all bands from the three spectral regions studied in this work in the
 form of stick diagrams.

\begin{table}[ht]
\caption{Comparison of calculated (TROVE) and experimental \cite{01MaBlSa.SO3} band centers and numbers
of line transitions.}
\centering
\begin{tabular}{rrrrrr}
\hline\hline
Band & Obs.  & Calc. & Pressure $P1^{a}$ & Pressure $P2^{b}$ & TROVE\\[0.5ex]
\hline
$\nu_2$ - $\nu_0$                   &     497.57   &   498.48     &        773   &       1265   &       5422     \\
$\nu_4$ - $\nu_0$                   &     530.09   &   528.59     &        996   &       2052   &      12195     \\
$\nu_1$ - $\nu_4$                   &     534.83   &     537.16   &          0   &         69   &      15147     \\
$\nu_2$ + $\nu_4$ - $\nu_2$         &     530.33   &     528.87   &         84   &        571   &      12477     \\
2$\nu_2$ - $\nu_2$                  &     497.45   &     496.88   &        112   &        704   &       7171     \\
$\nu_2$ + $\nu_4$ -$\nu_4$          &     497.81   &     498.76   &         47   &        602   &      27182     \\
2$\nu_4^{(l_4= 2)} - \nu_4$         &     530.36   &     528.79   &        116   &        775   &      31096     \\
2$\nu_4^{(l_4 = 0)} - \nu_4$        &     529.72   &     527.91   &         39   &        455   &      13718     \\
\hline 
$\nu_3$ - $\nu_0$                   &    1391.52   &    1387.45   &       2014   &    --        &      14441     \\
2$\nu_3^{(l_3 = 2)} - \nu_0$        &    2777.87   &    2770.29   &       1527   &    --        &      18115     \\
$\nu_0$ - $\nu_0$                   &    --        &     --       &         25   &    --        &       3439     \\
\hline\hline
\end{tabular}
\begin{tablenotes}
\item[a] For the $\nu_3$ and 2$\nu_3$ bands, measurements were taken at 0.16 Torr and 4.99 Torr respectively. Pressure values
are not recorded for microwave measurements \cite{91MeSuDr.SO3}. The remainder are the bands within the 405 - 708 \cm\ window, measured at 0.409 Torr
(Maki \etal~\cite{01MaBlSa.SO3}).
\item[b] Bands within the 405--708 \cm\ window, measured at 2.04 Torr (Maki \etal~\cite{01MaBlSa.SO3}).
\end{tablenotes}
\label{numLines}
\end{table}

\begin{table}
\caption{\label{bandInts}
Vibrational band intensities $S^{\rm exp}$, $S_{\rm tot}^{\rm calc}$, $S_{\rm red}^{\rm calc}$ in cm/molecule$\times 10^{-18}$,
and calculated transition moments $\bar\mu_{\rm if}$ in Debye.  $P_1$ and $P_2$ refer to the different pressure measurements within
 the same wavenumber region (see Table~\ref{numLines}). $S_{\rm red}^{\rm calc}/S^{\rm exp} $ is the ratio of the theoretical reduced
 and total band intensities (see text).  $S_{\rm tot}^{\rm calc}$ is the theoretical band intensity computed by summing all TROVE lines.
 $S^{\rm exp}$ is the experimental band intensity obtained from a summation over all experimental values after scaling factors applied (see text).
 $S_{\rm red}^{\rm calc}$ is the theoretical band intensity computed using only lines for which experimental counterparts exist. $N_{\rm red}$
 is the number of matched lines.}
\centering
\begin{tabular}{l cc c cc ccccc}
\hline\hline
Band     & \multicolumn{2}{c}{$S^{\rm exp}$}& $S_{\rm tot}^{\rm calc}$  & \multicolumn{2}{c}{$S_{\rm red}^{\rm calc}$}& \multicolumn{2}{c}{$S_{\rm red}^{\rm calc}/S^{\rm exp} $} & \multicolumn{2}{c}{$N_{\rm red}$} & $\bar\mu_{\rm if}/ D$    \\
\hline
                                    & $P_1$    & $P_2$    &           & $P_1$       & $P_2$     & $P_1$     & $P_2$     & $P_1$     & $P_2$    &         \\
\hline  
$\nu_2$                             &    2.987 &    1.537 &     3.705 &       2.915 &     1.559 &      0.98 &      1.01 &       773 &     1265 &    0.158\\
$\nu_4$                             &    4.258 &    3.112 &     5.949 &       4.310 &     3.149 &      1.01 &      1.01 &       995 &     2052 &    0.200\\
$2\nu_2$ - $\nu_2$                  &    0.116 &    0.470 &     0.661 &       0.101 &     0.411 &      0.88 &      0.88 &       112 &      704 &    0.221\\
$\nu_2 + \nu_4$ - $\nu_2$           &    0.062 &    0.322 &     0.528 &       0.052 &     0.251 &      0.84 &      0.78 &        84 &      571 &    0.199\\
$\nu_2 + \nu_4$ - $\nu_4$           &    0.026 &    0.260 &     0.581 &       0.022 &     0.215 &      0.84 &      0.83 &        47 &      602 &    0.223\\
$2\nu_4$($l_4$ = 2) - $\nu_4$       &    0.112 &    0.589 &     0.873 &       0.102 &     0.485 &      0.91 &      0.82 &       116 &      769 &    0.283\\
$2\nu_4$($l_4$ = 0) - $\nu_4$       &    0.026 &    0.222 &     0.405 &       0.015 &     0.179 &      0.57 &      0.81 &        38 &      454 &    0.196\\
$\nu_1$ - $\nu_4$                   &  --      &    0.009 &     0.101 &  --         &     0.003 & --        &      0.29 &  --       &       69 &    0.039\\
\hline 
$\nu_3$                             &   39.490 &  --      &    44.440 &      39.490 &   --      &  --       &  --       &      2014 &  --      &    0.321\\
$2\nu_3$                            &    0.093 &  --      &     0.119 &       0.093 &   --      &  --       &  --       &      1527 &  --      &    0.012\\
\hline\hline
\end{tabular}
\end{table}

\section{Conclusions and discussions}

The misplaced theoretical bands in Figures~\ref{allbands500} and \ref{so3fundamentals}
indicate that our \ai\ PES of SO\3\ requires improvement (see also band centers in Table~\ref{numLines}). The theoretical  $\nu_2$ frequencies
 have an rms deviation of 0.91 \cm\ when compared with the experimental data, which is relatively small compared to the deviation for $\nu_3$ of
 4.07 \cm. This is to be expected since our PES is purely \ai\ computed at a modest level of theory. We are planning to refine this surface by
 fitting to all experimental data available.

Table ~\ref{bandInts} outlines the quality of the intensity scaling procedure, in which the relative values of the experimental intensities where converted to absolute values (cm/molecule).
For the 0.409 Torr and 2.04 Torr measurements our comparisons mostly agree to within ~20\%, with the exception of the $2\nu_4^{(l_4 = 0)}$ - $\nu_4$ band measured at 0.409 Torr which shows nearly a 50\%\ difference, and the $\nu_1$ - $\nu_4$ band measured at 2.04 Torr with 80\%\ uncertainty. The latter can probably be attributed to both the small number of lines in each case, and residual errors in the transition dipole.
In the case of the $2\nu_4^{(l_4 = 0)}$ - $\nu_4$ band the comparison with the 2.04 Torr experiment yields a better value for $S_{\rm red}^{\rm calc}$/$S^{\rm exp}$ than for the 0.409 Torr measurement, which suggests that the number of lines available at 0.409 Torr  is too low (see also Figure~\ref{allbands500})
The significance of the  results  presented in Table~\ref{bandInts} and illustrated in Figure~\ref{allbands500} is that these give an estimation on the quality of the \ai\ dipole moment surface as well as of the experimental data. Based these numbers we can place a lower estimate on the quality  uncertainty for our intensities for each band, for example the experiment and theory for the $\nu_2$ and $\nu_4$ bands agree at least to within 3\% for both pressure measurements, while it is only 13\%\ for $2\nu_2$ - $\nu_2$, and between 17\%\ -18\%\ for the remaining bands.


Our complete room-temperature line list for SO\3\ containing 349~348~513 transitions can be accessed online at {\it{www.exomol.com}} in the ExoMol format described in \cite{12TeHiYu.exo}. It includes  the transition energies, Einstein coefficients $A(f \leftarrow i)$, and absorption intensities estimated for $T$ = 298.15~K.
Additionally, a list of 10~878 experimental transitions with absolute intensities obtained for $T$ = 296~K  is included into the supplementary part of this paper in a form suitable for standard atmospheric and planetary spectroscopic databases.


Our future work will be focused on the development of a hot line list for \sothree\ for high temperature industrial applications as well as for modelling molecular opacity in atmospheres of (exo-)planets  and cool stars as part of the ExoMol project \cite{jt528} (see {\it{www.exomol.com}}).

\section*{Acknowledgment}

We thank Alexander Fateev for stimulating our interest in this molecule
and for many helpful discussions, and Jeff Barber for supplying his
experimental result. This work was supported by grant 10442 from energinet.dk
under a subcontract from the Danish Technical University and
and the ERC under Advanced Investigator Project 267219.

\section{References}


\providecommand*{\mcitethebibliography}{\thebibliography}
\csname @ifundefined\endcsname{endmcitethebibliography}
{\let\endmcitethebibliography\endthebibliography}{}

\end{document}